\documentclass[extra,mreferee]{gji}
\usepackage{timet,color}
\usepackage{times}
\usepackage[urlcolor=blue,citecolor=black,linkcolor=black]{hyperref}
\usepackage{amsmath}
\usepackage{amsfonts}
\usepackage{amssymb}
\usepackage{fontenc}
\usepackage{graphicx}
\usepackage[utf8]{inputenc}

\title[$b$-value maps]
  {Richter $b$-value maps from local moments of seismicity}
\author[M. Holschneider et al.]
  {M. Holschneider$^1$\thanks{DFG, SFB 1294}, K. Ferrat$^1$, 
  G. Zöller$^1$\thanks{DFG, SFB 1294}, Ch. Molkenthin$^1$\thanks{DFG, SFB 1294}, S. Hainzl$^2$\\
  $^1$ University of Potsdam, $^2$GFZ Potsdam
  }
\pubyear{2020}

\begin{document}

\label{firstpage}

\maketitle

\begin{summary}
	We develop a technique to estimate spatially varying seismicity patterns. It is
	based on a Gaussian approximation of the underlying Poisson
	Process.  A link function is used to estimate local moments of the seismicity from
	observed catalogues. These are modeled by a nonstationary Gaussian field. 
	We construct a prior based on the local distribution of seismic faults. 
	This allows us to incorporate geological information into the Bayesian inversion 
	of the observed seismicity. In this paper we limit ourselve to the $b$-value 
	field for which we compute the  posterior expectations as well as the uncertainties. 
	The technique however may
	be applied to other seismically relevant parameters like Omori $c$ and $p$-values.
\end{summary}

\begin{keywords}
	$b$-value maps, non-stationary process of seismicity, Gaussian process analysis, correlation patterns
\end{keywords}

\section{Introduction}
Seismic hazard assessment relies on a reliable characterization of the frequency-magnitude distribution of the earthquakes. The general form of the distribution is usually well described by the Gutenberg-Richter (GR)  relation, the most famous statistical characteristic of earthquakes \citep{Gutenberg-Richter-1954}.
This widely accepted empirical formula, which has been validated numerous times by earthquake catalogs on a local and a global scale, states that the number $N$ of earthquakes with magnitude larger or equal to $m$ is 
\begin{equation}
  \log_{10} N (m) = a - b m , 
  \label{gr_law}
\end{equation} 
where magnitudes are assumed to be complete above a certain threshold $m_0$ depending on the earthquake catalog.
Here the $a$-value defines the total number of $m\ge 0$ events and $b$ the shape of the distribution. While the $a$-value largely varies depending on the region and the considered time and space selection, the $b$-value is found to be rather universal and scatters around one. Nevertheless strong local variations are reported with typical variations within the range $0.4<b<2.0$ \citep{Wiemer-Wyss-2002}.
Laboratory experiments have shown that the $b$-value describing the size distribution of acoustic emission events decreases with differential stress \citep{Scholz-1968,Amitrano-2003,Goebel-etal-2013} which seems to be in agreement with observations for earthquakes \citep{Schorlemmer-etal-2005,Narteau-etal-2009,Spada-etal-2013}. Therefore observed variations of the b-value in space and time are often interpreted as a stress meter, with an inverse relation between stress level and b-value \citep{Scholz-2015}. 

The most commonly used methods for mapping $b$-values in space are the fixed-radius and the nearest-neighbor methods \citep{Wiemer-Wyss-2002}. In the case of the fixed-radius approach, the $b$-value is calculated for earthquakes within a fixed radius and projected to the central point. In contrast, in the case of the nearest-neighbor method, the $b$-value of the grid-point is estimated from the $n$ closest events within a proximity limit. In a more recent work, \cite{Kamer-Hiemer-2015} incorporate a partitioning scheme based on Voronoi tessellation, where the number of random Voronoi nodes is determined by the Bayesian Information Criterion.

In this paper now we propose an alternative method, 
which is based on Gaussian process regression. We formulate it as correlation / covariance based method which has proved useful in the field of geomagnetism, where it has been named "correlation based modeling" \citep{Holschneider2016}

\section{Local moments of seismicity}

We suppose that in a region, the occurrence of earthquakes can be described
 through a non-stationary space, time, magnitude Poisson point process (ppp).
 In order to simplify the formulas, we consider the time stationary case only, but
 extension to non-stationary processes is immediate. Then, the Poisson intensity reads
 \begin{equation}\label{eq:grassumption}
	 \Lambda(x,m) = \alpha(x) \beta(x) e^{-\beta(x) (m-m_0)},\quad m\geq m_0,
 \end{equation}
 with $m_0$ the lower completeness bound for the magnitude $m$, the function 
$\alpha(x)$ is the local 
 seismic rate of events and $\beta(x)$ the local Gutenberg Richter 
 exponent, which is linked to the classical $b$-value through 
\[ 
 \beta(x) = \log(10) \ b(x).
 \]
 The likelihood of the Poisson point process generating an observed catalogue $\{ (x_i, m_i) \}$ is
\begin{equation*}
	\begin{aligned}
		&\log L ( (x_i, m_i) | \alpha(x) \beta(x) ) =\\
		 &\quad \sum_i \log \alpha(x_i) - \log \beta(x_i) - 
		 \beta(x_i) ( m_i - m_0) - \int \alpha (x) dx.
	\end{aligned}
\end{equation*}
A possible approach is to equip the fields $\alpha$ and $\beta$ with a suitable prior or penalizing structure $P(\alpha,\beta)$ for instance in form of a Gaussian process prior, and then obtain the posterior information via
\[
P(\alpha, \beta | ((x_i, m_i)) \sim L( (x_i, m_i) | \alpha(x) \beta(x) ) P(\alpha,\beta).
\]
A straightforward application of this approach is feasible but numerically demanding, see e.g. \citep{gu1993}.
In particular if not only the maximum a posteriori (MAP) estimator is to be deduced but also posterior 
uncertainties are of interest. 

We will use the southern California catalog of \citep{Hauksson2012}.  
The purpose of this paper is to propose an easily computable estimate 
 of the field $\beta(x)$ using techniques of Gaussian process regression or correlation based modeling (see e.g. \cite{}). As prior information we shall include 
 geological features like dominant fault directions (see 
 section~\ref{nonstationarypriorselection}). For this we shall use a Gaussian approximation of suitable moments that characterizes the
 distribution locally.  
 
 The first step is to realize that when the $\beta$ value in 
 the GR-law, which is a scaling parameter, becomes a translation parameter 
 on a log scale. More precisely, consider ($C$ is the Euler-Mascheroni constant)
 \[
 u = \log(m-m_0) + C,\quad C=0.577215\dots 
 \]
 and
 \[
 \nu =  - \log \beta.
 \]
 Then the density of $m$ (i.e. the Gutenberg Richter law) becomes
 \[
 \beta e^{-\beta m} dm = h(u - \nu) du,\quad h(u) = \exp(u-C - e^{u-C}). 
 \]
 The parameter $\nu$ is now the mean value of $u$
 \[
 \mathbb{E}(u|\nu) = \int u h(u-\nu) du = \nu. 
 \]
 and the variance of $u$ is the same for all $\nu$
 \[
\mathbb{V}(u|\nu) = \int u^2 h(u-\nu) - \nu^2 = \frac{\pi^2}{6}.
 \]
 This mean value can be estimated from local arithmetic means. In addition, the 
 uncertainties of these local means and their covariance properties may be
 estimated. This opens the way to a Gaussian posterior calculation. 
 
 In order to make this more precise, consider a small region $Q$ with center at $x$. 
 For a function $f(m)$ we define the local empirical moment as
 \[
 S_Q = \sum_{e_i \in Q} f(m_i) \quad
 M_Q = \frac{1}{N_Q} S_Q \quad N_Q = \#\{\mbox{events in $Q$}\}
 \]
 where the sum runs over all events $e_i=(x_i, m_i)$ in the region $Q$ and $N_Q$ is number of observed events in $Q$. 
 We assume that $N_Q >n\geq2$ for some fixed $n$. 
 We propose to use 
 \[
 u = f(m)=\log(m-m_0) + C, \quad C=0.577215\dots 
 \] so that $f$ takes values in all of $\mathbb{R}$. We therefore base our analysis on
 the following local seismicity moments
\begin{equation}\label{eq:logmagmoment}
	M_Q = \frac{1}{N_Q}\sum_{e_i \in Q} \log(m_i-m_0) + C.
\end{equation}
In order to assess the statistical properties of this quantity we transform the 
Poissonian intensity function to the field of 
  new point observables $(x_i, u_i)$, $u(x)=f(m(x))$. It then reads
 \[
\Lambda(x,u) = \alpha( x) h( u - \nu(x)),\quad \nu(x) = -\log(\beta(x)),
 \]
Here are the formulas that link the traditional $b$-value to $\beta$ and to our newly introduced 
 quantity $\nu$
 \[
 \begin{aligned}
     \nu(x) &=  - \log \beta(x) =  - \log \log 10 - \log b(x) \\ 
            &\simeq - 0.83403 - \log b(x) \\
            &\simeq   0.16597 - b(x)\quad\mbox{(for $b\simeq 1$)}\\
    \beta(x) &=  \log(10) b(x) \simeq 2.30258\ b(x)\\
            b(x)  &= \frac{e^{-\nu(x)}}{\log(10)} \simeq 0.4342944 e^{-\nu(x)}
 \end{aligned}
 \]

 We propose to perform a Bayesian inversion for the $\nu$ field. For this we choose the prior
 information about $\nu=-\log(\beta)$ to be  a Gaussian process described through prior mean
and prior covariance of the form
\[
\mathbb{E}(\nu(x)) = \mu(x), \quad \mathbb{V}(\nu(x), \nu(x^\prime)) = \Gamma(x,x^\prime)
\]
In section~\ref{nonstationarypriorselection} we will see, how to choose the covariance $\Gamma$ to take into account geological 
prior information about the seismicity generating structures. But even without this geological 
information, a Gaussian process description is useful for its flexibility and its direct computability. 
For the total seismicity $\alpha(x)$ we do not consider a random prior 
although this would be possible. We approximate the total seismicity in a small 
region $Q$ through $N_Q$ and we shall use
\begin{equation}
	\label{eq:alphaest}
\alpha(x) \simeq \frac{N_Q}{|Q|},\quad x \in Q.
\end{equation}

\begin{figure}
	 \centering
	 \includegraphics[width=0.9\linewidth]{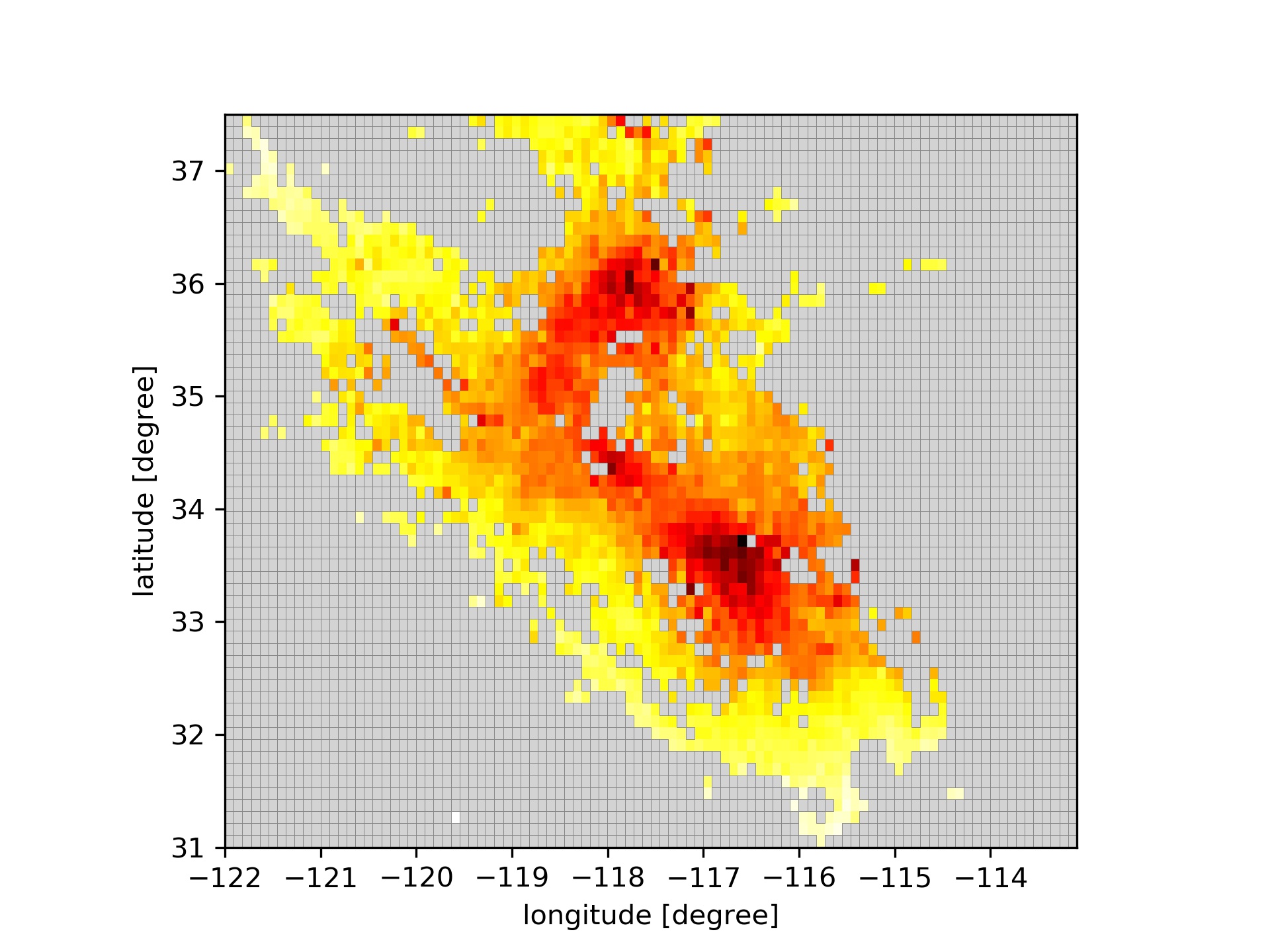}
	 \caption{The local moments of Eq.~\ref{eq:logmagmoment}. This is the surrogate 
	 data directly obtained from a catalogue. The size of each pixel patch is $10\times10\, km^2$. It serves as input to Gaussian process inversion yielding the Bayesian posterior distribution of $\nu=-\log(\beta)$.}
	 \label{fig:my_label}
 \end{figure}

 We now cover the observational region with rectangles $Q_i$, where we retain only those, 
 in which we have a minimal number of events. We take squares of $10\times10 km^2$ and we
 use only those bins, for which we have $N_i\geq 5$. 
 This leaves us with  $1968$ bins which comprise in total $593292$ events, whereas  $2253$
 events have been discarded. 
As shown in the appendix our 
 random model of seismicity the expectation of the local observational moment
 $S_i = S_{Q_i}$ for a realization of $\beta$ can be 
 computed (see Appendix, Eq.~\ref{eq:covarstructs}) 
\[
\mathbb{E}(S_i | \nu) = \int_{Q_i} \alpha(x) \nu(x) dx \simeq  \frac{N_i}{|Q_i|} \int_{Q_i} \nu(x).
\]
Actually, using local moments $M_i = S_i/N_i$, we measure the 
average  value of $\nu$ over $Q_i$ 
\[
\mathbb{E}(M_i| \nu) = \frac{1}{|Q_i|} \int_{Q_i} \nu(x).
\]
Under the random model of $\beta$ we have, as shown in the Appendix the following 
prior covariance properties for the local sum $S_i$
\[\begin{aligned}
	\mathbb{E}(S_i) &= \int_{Q_i} \alpha(x) \mu(x) dx\\
	\mathbb{V}(S_i) &= \int_{Q_i} \alpha(x) \Gamma(x,x^\prime) \alpha(x^\prime) +\\
	&\quad\int_{Q_i} \alpha(x) (\Gamma(x,x) + \mu(x)^2 + \frac{\pi^2}{6}) dx.\\
	\mathbb{V}(S_i, S_j) &= \int_{Q_i}\int_{Q_j} \alpha(x) \Gamma(x,x^\prime) \alpha(x^\prime) dx dx^\prime
\end{aligned}
\]
The prior covariance between the moments $S_i$ and the field $\nu$ is
\[
\mathbb{V}(\nu(x), S_i) = \int_{Q_i} \alpha(x^\prime) \Gamma(x, x^\prime) dx^\prime. 
\]
In general the local seismicity rate $\alpha$ is not known and must be estimated. We do not assume, that $Q_i$ is small since we will take into account 
the average of $\nu$ over $Q_i$ and not only the value of $\nu$ at the center of $Q_i$, which would be the "small $Q_i$"
approximation. So, provided the seismicity in $Q_i$ may be replaced with its average value, 
$Q_i$ may take any shape and size. 
In a forthcoming article, we will also remove the constraint on $\alpha$, by proposing a joint inversion of $\beta$ and $\alpha$. In this paper, 
however we take $\alpha$ as fixed, or at least locally estimated through the event count. 
Upon using 
the estimate 
\ref{eq:alphaest} we end up with the following correlation structure for the average moments $M_i$ ($i\not=j$)
\begin{equation}
	\label{eq:priorcovars}
\begin{aligned}
	\mathbb{E} (M_i)     &= \frac{1}{|Q_i|} \int_{Q_i} \mu(x) dx,\\
	\mathbb{V}(M_i) &= \Gamma_{i,i} = \frac{1}{|Q_i|^2}\int_{Q_i} \int_{Q_i}  \Gamma(x,x^\prime) dx dx^\prime +\\
	&\quad\frac{1}{N_i}\int_{Q_i} (\Gamma(x,x) + \mu(x)^2 + \frac{\pi^2}{6}) dx.\\
	\mathbb{V}(M_i, M_j) &=  \Gamma_{i, j} = \frac{1}{|Q_i||Q_j|}\int_{Q_i}\int_{Q_j} \Gamma(x,x^\prime) dx dx^\prime
\end{aligned}
\end{equation}
The first equation is used to compute the prior expectation of the observables (i.e. the moments).
The covariance between the observed moments and the field $\nu$ reads
\[
\mathbb{V}(\nu(x), M_{Q_i}) = \Gamma_{x,i} = \frac{1}{|Q_i|}\int_{Q_i} \Gamma(x,x^\prime) dx^\prime = \Gamma_{x, Q}.
\]
It is this covariance that allows us to infer on $\nu(x)$ from the observations of the $M_i$.
The precise formulation is given by Bayes theorem as exposed in the next section. 
All these quantities we need for the prior covariance structure 
may be computed a priori using numerical quadrature. 

\subsection{Bayesian inversion}

\begin{figure}\label{fig:posterior}
        \includegraphics[width=0.9\linewidth,]{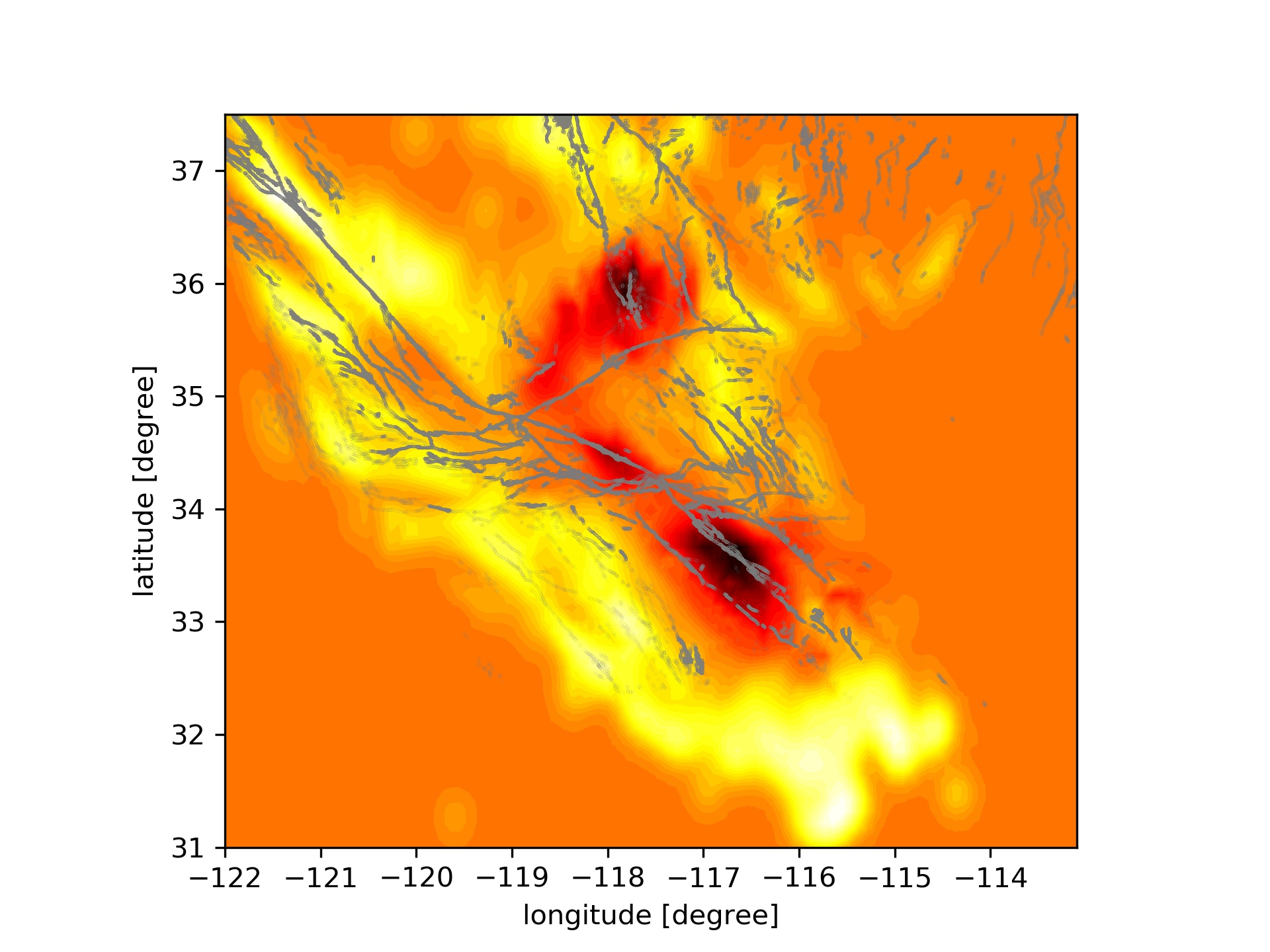}
	\caption{The posterior expectation of $\nu$. Dark regions correspond to high $\nu$, which corresponds to  small $b$-values. We observe how regions of high $\nu$ are localized near 
	the major faults.}
\end{figure} 

\begin{figure}\label{fig:posteriorvar}
        \includegraphics[width=0.9\linewidth,]{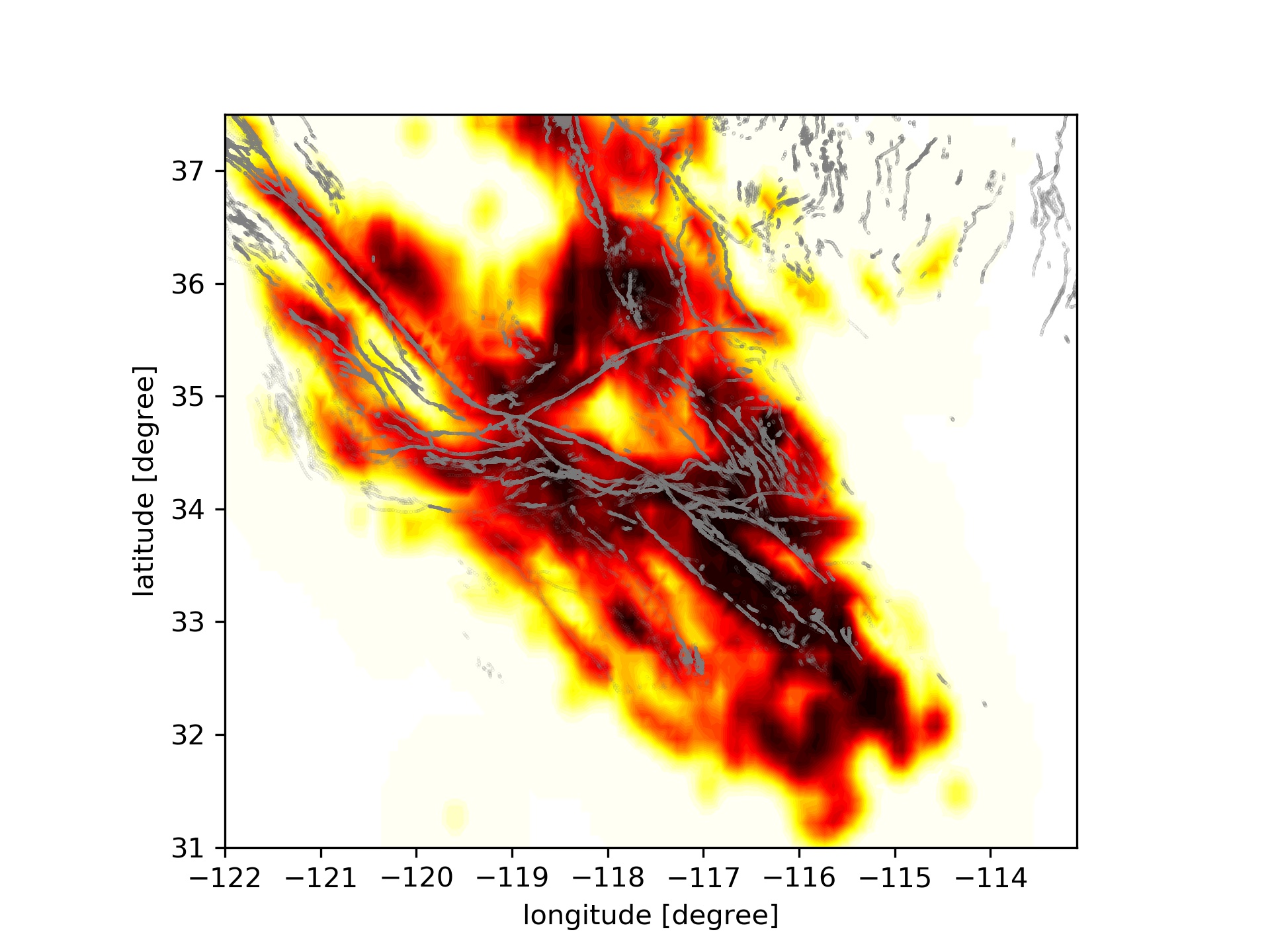}
	\caption{The diagonal part of the information gain \ref{eq:informationgain}.}
\end{figure}

After having computed all covariances, we may apply the Bayesian inversion 
for correlated Gaussian variables (see Appendix \ref{gaussianbayesiancalc}).
The key is the covariation of $\nu(x)$ and the $M_i$. 
In this setting the full posterior distribution of $\nu(x)$ 
given a collection of moment observations
$M_{Q_i}$ in disjoint regions $Q_i$, $i=1,\dots$ can be computed by standard 
Gaussian process calculus and we have
\[
\begin{aligned}
	&\mathbb{E}( \nu(x) | \{M_{Q_i}\} ) \\
	&= \mu(x) + \sum_k \Gamma_{x, Q_k} \sum_l \Gamma^{-1}_{k,l} 
	\left (M_{Q_l} - |Q_l|^{-1}\int_{Q_l} \mu(x) dx\right)
\end{aligned}
\]
The matrix $\Gamma$ with entries $\Gamma_{k,l}$
are the ones given by \ref{eq:priorcovars}.
We may also deduce the posterior uncertainty. It is described through a posterior covariance matrix  which reads
\begin{equation}
	\label{eq:informationgain}
\mathbb{V} (\nu(x), \nu(x^\prime) | \{M_{Q_i}\} ) =
\Gamma(x,x^\prime) - \sum_{k,l} \Gamma_{x, Q_k} \Gamma^{-1}_{k,l} \Gamma_{Q_l, x^\prime}.
\end{equation}
Observe that in principle the Bayesian inversion gives 
us a sub pixel image of the seismicity,  since we predict point
values of $\nu(x)=-\log \beta(x)$ from averages of seismicity in the regions $Q_i$. 
Clearly however, the prior covariance will influence the posterior blurriness of the inversion. 

\section{Non stationary prior selection}
\label{nonstationarypriorselection}

In the Bayesian inversion, we have incorporated as prior information 
the geometry of the fault system. For this we need a prior correlation, that reflects the
underlying fault geometry. 
A dominant geological feature of a seismogenic zone is the non isotropy of the fault directions. 
We propose to take this 
feature into account for the Bayesian inversion for 
the $\beta$ field in the following way. 

\begin{figure}\label{fig:faults}   
	\centering
       \includegraphics[width=0.9\linewidth,]{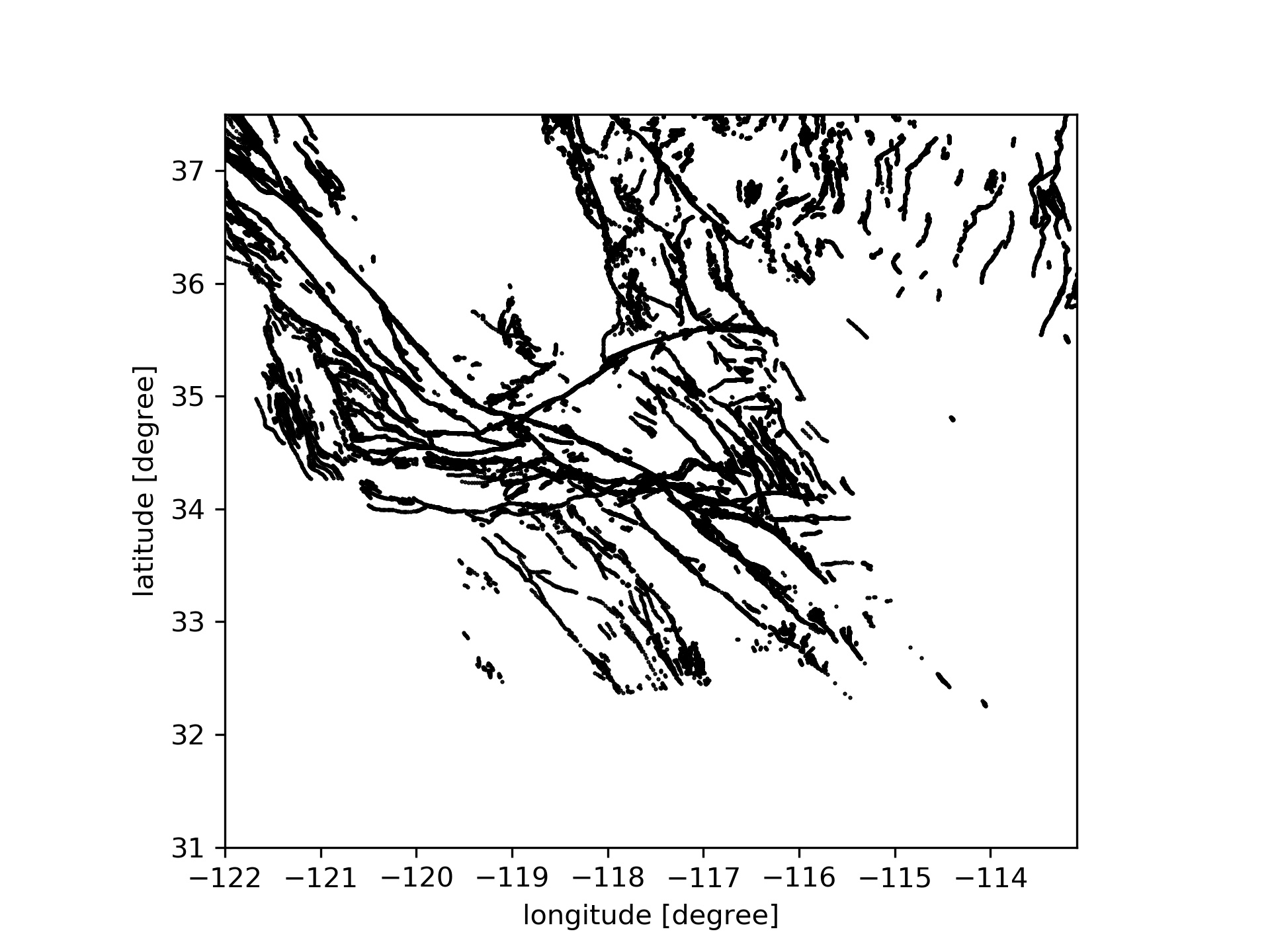}
	\caption{The surface expression of faults of southern California as found in "U.S. Geological Survey, 2006, Quaternary fault and fold database for the United States, accessed 08/08/2011, from USGS web site: http://earthquakes.usgs.gov/regional/qfaults/".}
 \end{figure}

\begin{figure}\label{fig:faultdirections}
	 \centering
        \includegraphics[width=0.9\linewidth,]{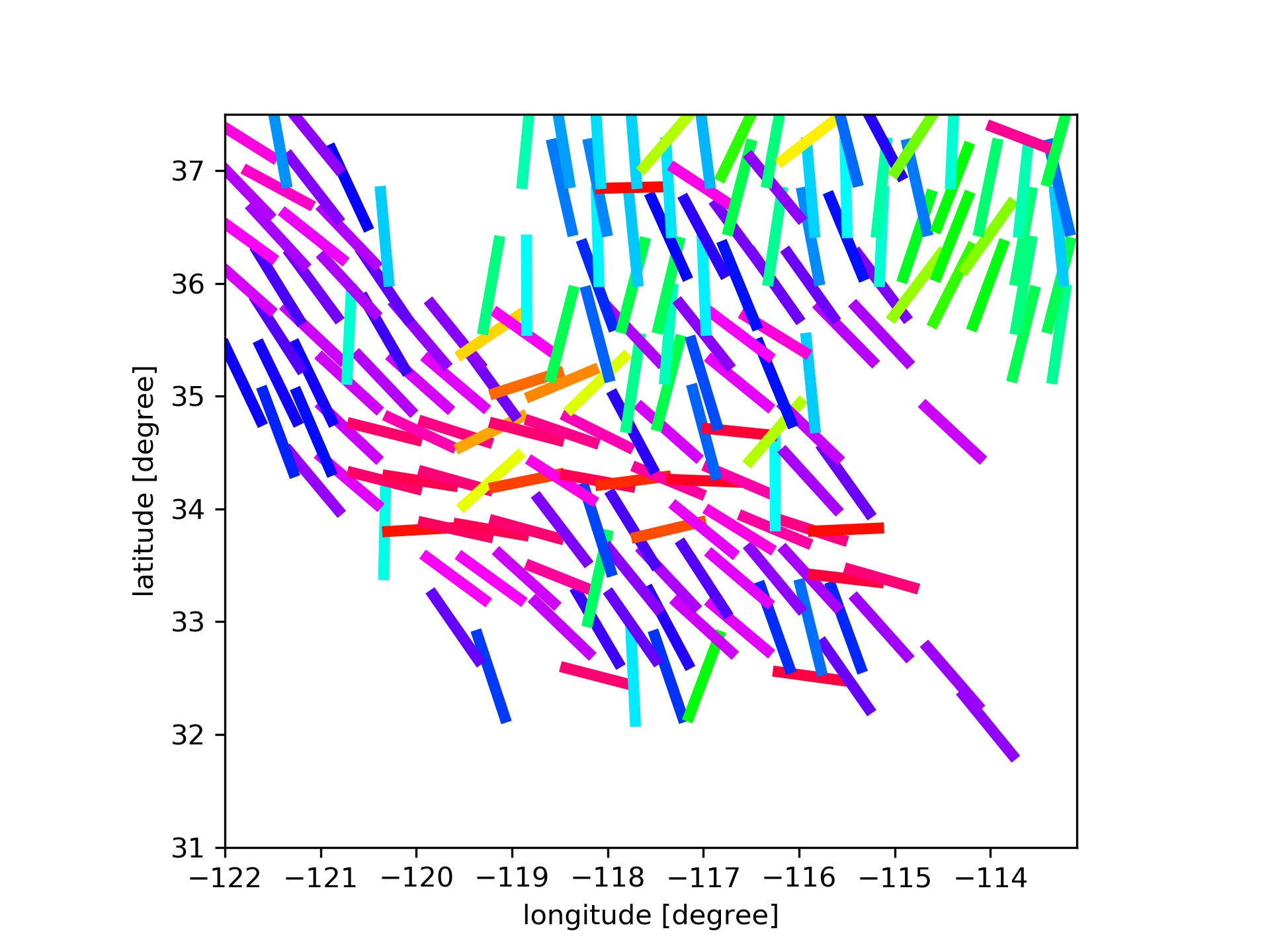}
	 \caption{The major direction of the faults. This is used as a basis for the directional 
	 covariance prior structure.}
\end{figure}

In a first step we determine at each point the dominant fault direction. For this we coarse grained the 
entire region into patches of $40\times 40 km$. In each patch we applied a median filter 
on the directions 
of the local fault segments in that region as they are present in the 
 data set (U.S. Geological Survey, 2006, Quaternary fault and fold database for the United States, accessed 08/08/2011, from USGS web site: http://earthquakes.usgs.gov/regional/qfaults/). 
 This is arguably not a very refined method to estimate the local main direction of 
expected correlation from geological data. However this paper is not concerned with fully exploiting the
possibilities such prior choices offer. We merely want to show, how in principle such information can be taken into account
in an estimation of a non-homogeneous $b$-field. 
In Fig.~\ref{fig:faultdirections} you can see the direction field obtained that way. 
 
This direction field is then used to construct a prior covariance kernel $\Gamma(x,x^\prime)$ in the following way. 
We suppose that along the main fault direction the a priori correlation length is $\sigma_1=20km$ whereas 
orthogonal to it we only assume a correlation length of $\sigma_2=5km$. If there are no faults in a $40km$ distance,
we set the prior correlation length to $\sigma_1 = \sigma_2 = 10km$. 
This gives rise to a covariance matrix 
\[
\Sigma_e = R_e^t \mbox{diag}(\sigma_1^2, \sigma_2^2) R_e.
\]
This defines a bell shaped correlation kernel
\[
	g(e, x)  \sim  \exp\ - \frac{1}{2}\left( \frac{(e^t x)^2}{\sigma_1^2} + \frac{x^t x - (e^tx)^2}{\sigma_2^2}\right)
\]
There is an implicit amplitude that is not written but shall be picked later. 
Now consider uncorrelated standard white noise
\[
\mathbb{E} (\omega(x)) = 0,\quad
\mathbb{V} (\omega(x), \omega(x^\prime)) = \delta(x-x^\prime).
\]
Upon convolving $\omega$ with the kernel $g$ in a non-stationary way, taking at each point the local dominant fault 
direction, we may define a random field that reflects the non-isotropic geology.
\[
s(x) = \int g(e_x, x-x^{\prime\prime}) \omega(x^{\prime\prime}) dx^{\prime\prime}.
\]
This is a zero mean random field and the covariance is (upon exchanging expectation and integration and using $\mathbb{E}\omega(t)\omega(t^\prime) = \delta(t-t^\prime)$)
\[
	\mathbb{V}(s(x), s(x^\prime) = \int g(e_x, x- x^{\prime\prime} ) g(e_{x^\prime}, x^\prime - x^{\prime\prime}) dx^{\prime\prime} 
\]
which by standard rules for the integration of Gaussians (see Appendix \ref{gauscalc} reads
\[
\begin{aligned}
	&\mathbb{V}(s(x), s(x^\prime) = \Gamma(x,x^\prime)\\
	&=\frac{\sigma^2|\Sigma_{x}|^{1/4}|\Sigma_{x^\prime}|^{1/4}}{| \Sigma_{x}+\Sigma_{x^\prime}|^{1/2}} 
	\exp - \frac{1}{2} (x-x^\prime)^t (\Sigma_{x}+\Sigma_{x^\prime})^{-1} (x-x^\prime). 
\end{aligned}
\] 
The  amplitude $\sigma^2$ is chosen to reflect the a priori uncertainty of $\nu=-\log(\beta)$. Since we expect $b$-values to variate about 20\% we have 
set $\sigma^2=0.4$.

\begin{figure}
	\label{fig:priorcovar} 
    \centering
        \includegraphics[width=0.9\linewidth,]{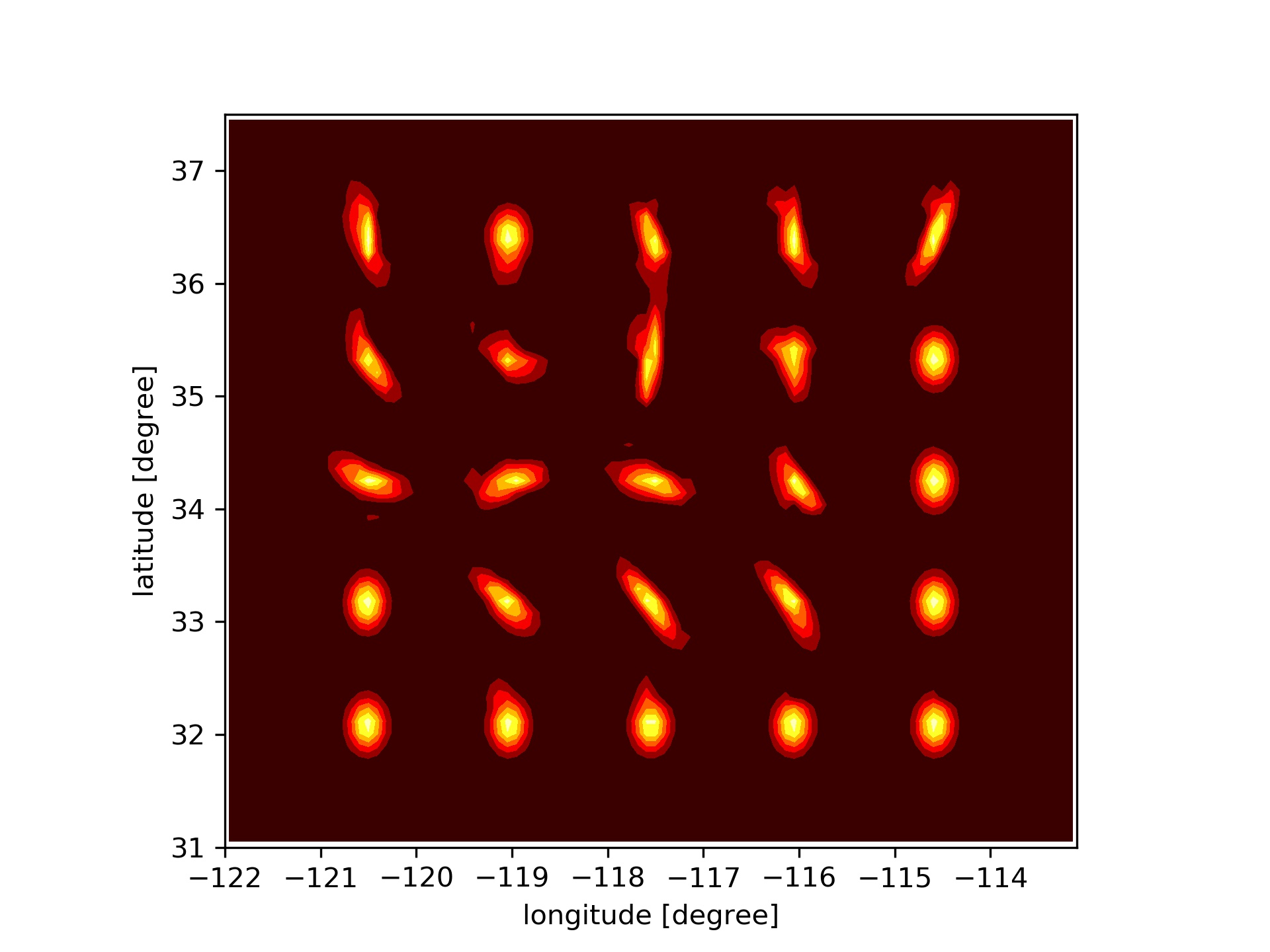}
	\caption{The prior correlation kernel for the $\nu =  - \log \beta$ field. It is based on the local dominant direction of the fault. The correlation structure corresponds to non-stationarily filtered white noise. We show a superposition of 
	$\Gamma(x,x_i)$ for $x_i$ taken in a grid.}
\end{figure}

In Fig.~\ref{fig:priorcovar} you can see the prior covariance kernel we are using. 
It clearly shows that the correlation tends to be oriented along the dominant fault 
directions.

At this stage this is not an exhaustive analysis of the possibilities this method opens, but it
is a feasibility study that shows, how in principle geological or other 
prior information may be build into
the estimation of the space (time) $b$ respectively $\nu$-value distribution.

\section{Conclusion}

We have introduced a Bayesian inversion of catalog data for $b$-value maps. It is based on the observation of local moments, 
which are approximated by local averages of magnitudes. The $b$-value field has been transformed using a $\log$ as link function
into a field for which a Gaussian process prior is assumed to hold. The covariance function could then incorporate 
additional information about the seismicity patterns. As an example, we have used local direction of faultings to include
preferred correlation directions into the prior covariance patter. The statistics of the local averages has been discussed in great detail, paying attention to fluctuations which come from the uncertainty of the field as well as of the Poissonian nature of
catalog generating filed. For this we have developed an extension of Campbell's sampling formula for random Poisson processes. 
As an illustration we have applied this technique to the estimation of 
$\nu = C -\log(\beta)$ value maps in 
southern California. 

\section{Appendix}

In this appendix we have collected the mathematical tools needed for the analysis
exposed in the paper. We have decided to separate this technical part from the rest of the
paper for the convenience of the reader. 

\subsection{Likelihood of Poisson point process}

Let $\Lambda(x)$ be the Poissonian intensity of a point process over a domain $\Omega$.
The the likelihood to observe a random catalog $x_1, x_2 \dots x_N$ is 
\[
\log  L( N, x_1, x_2 \dots x_N | \Lambda )  = \sum_i \log\Lambda(x_i) - \int \Lambda(x) dx + \mbox{Const}.
\]
Recall that $N$ is a  Poissonian random variable with frequency $\lambda = \int_\Omega \Lambda$
and given $N$, the points $x_1$, $x_2\dots x_N$ are uniformly distribution with density $\Lambda(x) / \lambda$.

\subsection{Gaussian Bayesian calculus}
\label{gaussianbayesiancalc}
Let $X$ and $Y$ be two jointly Gaussian random vectors with prior mean $\mu_X$, $\mu_Y$ and covariances
$\Gamma_{XX}$, $\Gamma_{YY}$ and $\Gamma_{XY}=\Gamma_{YX}^T$. Then the posterior information about $X$
that can be inferred from observing $Y$ is again Gaussian. The posterior mean value is
\[
\mathbb{E}(X|Y) = \mu_X + \Gamma_{XY} \Gamma_{YY}^{-1} ( Y - \mu_Y)
\]
and the posterior variance is
\[
\mathbb{V}(X|Y) = \Gamma_{XX} - \Gamma_{XY} \Gamma_{YY}^{-1} \Gamma_{YX}.
\]

\subsection{Integrals of Gaussians}
\label{gauscalc}
For a symmetric $2\times2$ matrix $\Sigma$ and $\mu\in\mathbb{R}^2$ 
consider
\[
g_{\mu,\Sigma} (x) = \frac{1}{2\pi|\Sigma|^{1/2}}e^{- (x-\mu)\Sigma^{-1}(x-\mu)/2}
\]
This function satisfies
\[
\int g_{\mu,\Sigma} (x) dx =1
\]
Since this is the pdf of a two dimensional Gaussian rdv with mean $\mu$ and variance $\Sigma$ we 
have, upon considering the sum of two such rdvs (recall that the pdf of the sum of two independent rdvs is their convolution)
\[
g_{\mu,\Sigma}\ast g_{\mu^\prime,\Sigma^\prime} = g_{\mu+\mu^\prime, \Sigma + \Sigma^\prime}
\]
and thus upon evaluation this expression at $x=0$ and using 
$g_{\mu,\Sigma}(-x)=g_{-\mu, \Sigma}(x)$
\[
\int g_{\mu,\Sigma} (x) g_{\mu^\prime,\Sigma^\prime} (x) dx = 
g_{\mu-\mu^\prime, \Sigma+\Sigma^\prime}.
\]
Now consider filtered two dimensional white noise $\omega = dW$
\[
s(x) = 2\sqrt{\pi} \gamma(x) |\Sigma_x|^{1/4} \int g_{x, \Sigma_x}(x^\prime) dW(x^\prime),
\]
with some arbitrary modulating amplitude $\gamma(x)$. This is a zero mean Gaussian random process 
and we have for the covariance
\[\begin{aligned}
	&\mathbb{V}(s(x), s(y)) \\
	&= 2 \gamma(x) \gamma(y) \frac{|\Sigma_x|^{1/4}  |\Sigma_y|^{1/4}}{|\Sigma_x + \Sigma_y|^{1/2}} \\
	& \quad \times 
\exp - (x-y)^T (\Sigma_x + \Sigma_y)^{-1}(x-y)/2.
\end{aligned}
\]
In particular, setting $x=y$ we see that  the variance at a point $x$ is 
precisely $\gamma(x)^2$
\[
\mathbb{V}(s(x) ) = \gamma(x)^2.
\]
And the correlation structure changes locally with the chosen 
matrix valued function $x\mapsto \Sigma_x$.

\subsection{Total mean and total variance}

A basic tool that we will use is the formula of total mean
\[
\mathbb{E}(A) = \mathbb{E}(\mathbb{E}(A|C))
\]
and of total covariance
\[
\mathbb{V} (A,B) = \mathbb{E}( \mathbb{V}(A,B|C) ) + \mathbb{V}( \mathbb{E}( A|C)
\mathbb{E}( B|C)) 
\]
In particular we have for the variance
\[
\mathbb{V} (A) = \mathbb{E}( \mathbb{V}(A|C) ) + \mathbb{V}( \mathbb{E}( A|C)).
\]

\subsection{A random Campbell theorem}

For a function $f$ and random points from a Poisson point process with intensity function $\Lambda(x)$ consider the  random sums of the form
\[
S = \sum f(x_i)
\]
Then Campbell's theorem states that the expectation reads
\[
\mathbb{E} (S) = \int f(x) \Lambda(x) dx
\]
and that the variance is
\[
\mathbb{V} (S) = \int ( f(x) )^2 \Lambda(x) dx.
\]
An extension of this result for two functions $f$, $g$
yields the following covariance
\[
\mathbb{V} (S_f, S_g) = \int f(x) g(x)  \Lambda(x) dx.
\]
We need an analogue theorem, but where the density function $\Lambda$ itself is a random process. 

We consider only the second order statistics of $\Lambda$
\[
\begin{aligned}
	\mu(x) &= \mathbb{E}(\Lambda(x)),\\ 
	\Gamma(x,x^\prime) &= \mathbb{V}(\Lambda(x), \Lambda(x^\prime)) \\ &= 
	\mathbb{E}\left((\Lambda(x) - \mu(x))(\Lambda(x^\prime) - \mu(x^\prime)))\right).
\end{aligned}
\]
Then Campbell's theorem gives us $\mathbb{E}(S|\Lambda)$ and $\mathbb{V}(S|\Lambda)$.
The direct application of the total variance theorem yields
\[
\begin{aligned}
	\mathbb{E}(S_f) &= \int f(x) \mu(x) dx,\\
	\mathbb{V}(S_f, S_g) &= \int f(x)g(x) \mu(x) dx + \int f(x) \Gamma(x,x^\prime) g(x^\prime) dx dx^\prime
\end{aligned}
\]
We have used the following useful formulas that apply when working with second order properties of the processes $\Lambda$
\[
\begin{aligned}
	\mathbb{E} \int f \Lambda &= \int f  \mu,\\
	\mathbb{E} \int f  \Lambda \int g \Lambda &= \int f \Gamma g + \int f  \mu  \int f  \mu \\
	\mathbb{V} \left(\int f  \Lambda, \int g \Lambda \right )&= \int f  \Gamma g
\end{aligned}
\]
We also need the covariance between a linear functional of $\Lambda$ and the moments $S_f$. 
So let 
\[
L = \int l(x) \Lambda(x) dx.
\]
Then, for fixed $\Lambda$ the quantity $L$ and $S$ are independent, and thus using 
the total variance calculation we obtain
\[
\begin{aligned}
	\mathbb{E} (L) &= \int l(x) \mu(x) dx\\
	\mathbb{V} (L S_f)&= \int \!\!\int l(x) \Gamma(x,x^\prime) f(x^\prime) dx dx^\prime
\end{aligned}
\]

\subsection{A hierarchical (marked) family of random intensities}

Consider the following hierarchical (or marked) setting. 
We suppose that $x$ actually has two parts $(x,u)$ and that
$\Lambda$ factorises as follows
\[
\Lambda(x,u) = \mathbb{P}(u | x) \alpha(x).
\]
For the conditional distribution we assume that with some probability density function $h$ we have
\[
\mathbb{P}(u|x) = h( u - \nu(x) ),
\]
with some random function $\nu$. So, a realization of events from $\Lambda$ are sampled as follows: 
The function $w$ is chosen and then fixed. We now 
pick $x$ at random from $\alpha(x)$ and then 
$u$ is picked at random for each $x$ according to the density $h(u-\nu(x))$
\[
\Lambda(x,u) = \alpha(x) h( u - \nu(x) )
\]
Without loss of generality we may assume that the density function $h$ is such that $u\sim h$ has
zero expectation. Then $h$ satisfies
\[
\int h(u) = 1, \quad \int u h(u) du = 0, \quad \int u^2 h(u) =  \sigma^2.  
\]
For the function $\nu(x)$ we assume that it is a realization of 
a Gaussian process with mean and covariance structure given by
\[
\begin{aligned}
	\mathbb{E}(\nu(x)) &= \mu(x),\\ 
	 \mathbb{V}(\nu(x), \nu(x^\prime)) &= \Gamma(x,x^\prime). 
\end{aligned}
\]
Let $Q$ be a region and let $\chi_Q(x)$ be its indicator function. Then we may apply 
the general theory of the previous section 
to the function 
\[
f(x,u)= \chi_Q(x) u.
\] 
The local sample moments are then
\[
S_Q = \sum_{x_i\in Q} u_i.
\]

We now compute all the moments and their second order statistics for this random family.
The $f$ moment for fixed $\nu$ reads
\[
\int f(x,u) \Lambda(x,u) dx du = \int_Q\int u \Lambda(x,u) du dx = \int_Q \alpha (x) v(x) dx,
\]
which implies that for the moment of two point covariance we can write 
\[\begin{aligned}
	\int f(x,u) &\Gamma(x,u, x^\prime, u^\prime) f(x^\prime,u^\prime) = \mathbb{V}(\int f \Lambda)\\
	& = \int_Q\int_Q \alpha_Q(x) \Gamma(x,x^\prime) \alpha_Q(x^\prime) dx dx^\prime.
\end{aligned}
\]
In the same way we may write 
\[
\begin{aligned}
	&\int_Q \int f(x,u)^2 \mu(x,u) du dx \\ &\quad = \mathbb{E} \int_Q \int u^2 \hat\alpha(x) h(u-\nu(x) ) du dx \\  
	&\quad = \sigma^2 \int_Q \alpha(x) dx  +
	\int_Q \alpha(x) (\Gamma(x,x) + \mu(x)^2) dx
\end{aligned}
\]
For two disjoint regions $Q$ and $Q^\prime$ we have $\int \mu(x) f_Q(x) f_{Q^\prime}(x) dx =0 $. 

To summarize we have the following second order statistics for the local moments $S_Q$ and $S_{Q^\prime}$ in two disjoint regions $Q$ and $Q^\prime$
\begin{equation}
\label{eq:covarstructs}
\begin{aligned}
	\mathbb{E}(S_Q) &= \int_Q \alpha(x) \mu(x) dx.\\
	\mathbb{V}(S_Q) &= \int_Q\int_Q \alpha(x) \Gamma(x,x^\prime) \alpha(x^\prime) dx dx^\prime\\
		&\quad + 
	\int_Q \alpha(x) (\Gamma(x,x) + \mu(x)^2 + \sigma^2) dx \\
	\mathbb{V}(S_Q, S_{Q^\prime}) &=  \int_Q \int_{Q^\prime} \alpha(x) \Gamma(x,x^\prime) \alpha(x^\prime) dx dx^\prime
\end{aligned}
\end{equation}
Consider now the functional 
\[
l_y(x,u) = \frac{1}{\alpha(x)}\delta(y-x) u,
\]
with $\delta$ the Dirac unit mass function. It acts as measuring $\nu$ 
at the point $y$ since
\[
L_y = \int l_y(x,u) \Lambda(x,u) dx du =  \nu(y)
\]
Its covariance with the $S_Q$ is the covariance between $\int l_y\Lambda = \nu(u)$
and $\int f_Q \Lambda = \int_Q \alpha \nu$, which is
\[
\mathbb{V}(L_y, S_Q) = \mathbb{V}(\nu(y), S_Q) = \int_Q \alpha(x) \Gamma(y,x) dx. 
\]

\bibliographystyle{gji}
\bibliography{bibliography}

\label{lastpage}

\end{document}